\RequirePackage{fix-cm}
\documentclass[singlecolumn,epjc3]{svjour3}
\smartqed  
\RequirePackage{graphicx}
\usepackage{graphicx}
\usepackage{graphicx,amsfonts,amssymb}
\usepackage{amsmath,amssymb,enumerate}
\usepackage{placeins}
\usepackage{multirow}
\usepackage{multicol}
\usepackage{float}
\journalname{Eur. Phys. J. C}

\begin{document}
\title{Generalized relativistic anisotropic models for compact stars}
	
	\author{S.K. Maurya\thanksref{e1,addr1}
		\and Y.K. Gupta\thanksref{e2,addr2}
		\and Baiju Dayanandan \thanksref{e3,addr1}
		\and M. K. Jasim \thanksref{e4,addr1}\\
		\and Ahmed Al-Jamel \thanksref{e5,addr1,addr3}.}
	
	\thankstext{e1}{e-mail: sunil@unizwa.edu.om}
	\thankstext{e2}{e-mail: kumar$001947$@gmail.com}
	\thankstext{e3}{e-mail: baiju@unizwa.edu.om}
	\thankstext{e4}{e-mail: mahmoodkhalid@unizwa.edu.om}
	\thankstext{e5}{e-mail: aaljamel@unizwa.edu.om}
	
	\institute{Department of Mathematical \& Physical Sciences, College of Arts \& Science,
		University of Nizwa, Nizwa, Sultanate of Oman\label{addr1}
		\and Department of Mathematics, Raj Kumar Goel Institute of Technology, Ghaziabad, 201003, U.P. (INDIA) \label{addr2}
		\and Department of Physics, Al Al-Bayt University, Mafraq 25113, Jordan\label{addr3}}
		
	\date{Received: date / Accepted: date}
	
	\maketitle
	
\begin{abstract}
	we present new anisotropic generalization of Buchdahl \cite{1} type perfect fluid solution by using the method of earlier work \cite{2}. In similar approach we have constructed the new pressure anisotropy factor $\Delta $ by the help both the metric potential $e^{\lambda } $ and $e^{\nu } $. The metric potential $e^{\lambda } $ same as Buchdahl [1] and $e^{\nu} $ is monotonic increasing function as suggested by Lake [3]. After that we obtain new well behaved general solution for anisotropic fluid distribution. We calculated the physical quantities like energy density, radial and tangential pressures, velocity of sound and red-shift etc.  We observe that these quantities are positive and finite inside the compact star. Also note that mass and radius of our models can represent the structure of realistic astrophysical objects such as Her X-1 and RXJ 1856-37.
	\end{abstract}


\section{Introduction}

The general theory of relativity (i.e. Einstein's theory of gravitation and space time) is the most beautiful and elegant of physical theories which made foundation for our understanding of compact relativistic astrophysical objects. The predictions of general relativity have been authenticated to be in harmony with observational data in relativistic astrophysics and cosmology. Numerous static perfect fluid compact star models of have been constructed by the help of exact solutions of Einstein's filed equations for static spherically symmetric line element in the past years because this is first approximation of building a realistic star models. There are a large number of models in literatures based on the topic of the spherically symmetric exact solutions but very few of them satisfy the required general physical situations inside the stellar interior dense matter \cite{4}. In this connection, The number of workers [5-11] have been investigated the structure of compact astrophysical object such as X-ray pulsar, Her X-1, RXJ 1856-37, X-ray busters, 4U 1820-30, SAX J 1808.4-3658, X-ray sources, 4U 1728-34 and PSR 0943+10, whose mass-radius relationship of this astrophysical objects are important because these can be directly compared with observational compact stars. The mass to radius relationship of the stellar object provide a vital clue to distinguish between different super dense stars, white dwarf, neutron stars and ultra-compact stars from one another.

The study of compact stellar structure and evolution is that the interior structure of a star can be modeled as anisotropic perfect fluid. Starting point in the studies of compact relativistic astrophysical objects is represented by the interior Schwarzschild solution. The works of Ruderman \cite{12} and Canuto \cite{13} on compact star having matter distributions with energy densities much greater than the nuclear regime indicate that the super-dense stars are likely to develop anisotropic pressure. It has been discovered that the anisotropic pressure is diverse in physical nature i.e. radial pressure is not equal to the tangential pressure inside the core. The issue of local anisotropy in pressure was extensively~reviewed by Herrera and Santos \cite{14} in a general relativistic approach and present several physical mechanisms for its origin in the systems of extremely low and very high density, which may include astrophysical compact objects. It would be expected that origin of pressure anisotropy~is due to a number of physical phenomena that may take place inside the gravitational compact objects. Bowers and Liang \cite{15} shown that~the~possible importance of locally anisotropic~equation of state in high density ranges with order  ${10}^{15}g{cm}^{-3}$~where nuclear reaction have to be treated relativist ally. It has been shown that due to geometry of~modes, anisotropic distribution of pressure could be considered to pion condensation or super fluidity state.~Also they have investigated that the anisotropy may also affect on the limiting values of the maximum mass of compact objects. Hernandez and Nuez \cite{16} obtained a general method for static spherically symmetric anisotropic~solutions which are obeying nonlocal equation of state form density profile by assuming the conditions of a vanishing Weyl tensor. Mak \& Harko \cite{17} found a class of exact solutions of Einstein's field equations characterizing spherically symmetric and static anisotropic matter distribution. Sharma and Maharaj \cite{18} have given a class of exact solution which can be applied to strange star with quark matter for neutral anisotropic matter.

In the recent years there have been several~investigations of the Einstein field equations corresponding to anisotropic matter distribution where the component of radial pressure differs from the angular component and examined how anisotropic matter distribution affects on the effective mass, radius of the stars, central energy density, critical surface red-shift and stability of highly compact bodies. These works can be seen in further references [19-28].

Now we would like to focus on our present work, As very recently Gupta and Maurya \cite{29} have obtained a class of well-behaved charged analogues of Buchdahl's \cite{1} neutral perfect fluid solution, which reduces to its natural counterpart in the absence of charge under the Einstein-Maxwell space times. The members of this class have been shown to satisfy various physical conditions. This analysis of the model reveals both vela and crab pulsars. In the present paper, we proceed in similar fashion to obtain the general anisotropic solution of Buchdahl's \cite{1} metric for compact star models.  Also this paper is sequel of the papers \cite{2,30}, where in paper \cite{30} we developed the general algorithm for charged anisotropic solutions for spherically symmetric metric, however in paper [2] we have adopted a different approach to construct the anisotropy factor with the help of both metric potential $e^{\nu } $  and  $e^{\lambda } $. The later one is very interesting method to construct the anisotropy factor by the help of the metric potentials.  As per as our knowledge and survey of literatures, till now no alternative well behaved anisotropic solution for compact star is available in the literature. Our main goal have twofold: first to find new exact anisotropic solution by the same methodology as in \cite{2} which may gives the better realistic model to express the physical properties of compact star models, secondly, to investigate the role pressure anisotropy on the maximum mass of compact stars.

The structure of the paper as follows: In Sec. 2 we set up the Einstein field equations for anisotropic compact objects are given whereas the method to construct the anisotropy and general solutions is shown in Sec. 3. In Sec. 4 we represent fascinating characteristics of the physical parameters which include central density, radial and tangential pressures, stability, pressure anisotropy, adiabatic index and red shift etc. along with the matching condition. We provide the effective- mass relation and red shift of the compact star in Sec.5. As a special study, In Sec.6 we compared our models to the observational compact star models with the several data sheets in connection to compact stars. At last we discussed the brief summary of discussions and conclusions. Hopefully, our results will be more important to analysis of structure of observational compact stars objects, as well as study of the behavior of matter undergoing strong gravitational field.

\section{Metric and The Einstein Field Equations}

We consider the static spherically symmetric metric to describe the anisotropic fluid distribution with Schwarzschild coordinates,~$x_{i} =(r,\,\theta ,\, \phi ,\, t)$ \cite{29,30}
\begin{equation}
ds^{2} =-e^{\lambda } dr^{2} -r^{2} (d\theta ^{2} +\sin ^{2} \theta \, d\phi ^{2} )+e^{\nu } \, dt^{2}
\label{1}
\end{equation}
We assume the Einstein field equations for a static spherically symmetric anisotropic fluid distribution in the standard form of ordinary differential equations \cite{31},
\begin{equation}
\label{2}
\kappa \, p_{r} =\frac{v'}{r} e^{-\lambda } -\frac{(1-e^{-\lambda } )}{r^{2} }   ,
\end{equation}
\begin{equation}
\label{3}
\kappa \, p_{t} =\left[\frac{v''}{2} -\frac{\lambda 'v'}{4} +\frac{v'^{2} }{4} +\frac{v'-\lambda '}{2r} \right]\, e^{-\lambda } ,
\end{equation}
\begin{equation}
\label{4}
\kappa \, c^{2} \rho =\frac{\lambda '}{r} e^{-\lambda } +\frac{(1-e^{-\lambda } )}{r^{2} }
\end{equation}
\textbf{Pressure anisotropy condition: }
\begin{equation}
\label{5}
\Delta =\kappa \, (p_{t} -\, p_{r} )=\left[\frac{v''}{2} -\frac{\lambda 'v'}{4} +\frac{v'^{2} }{4} -\frac{v'+\lambda '}{2r} -\frac{1}{r^{2} } \right]\, e^{-\lambda } +\frac{1}{r^{2} } ;
\end{equation}
Now let us take the metric potential according the Buchdhal \cite{1} space time:
\begin{equation}
\label{6}
e^{-\lambda } =\frac{2-Cr^{2} }{2\, (1+Cr^{2} )} ;   \nu =2\, \log \, y,
\end{equation}
Eq.(\ref{5}) becomes:
\begin{equation}
\label{7}
\Delta =\,\left[ \frac{y''}{y} -\frac{2+4Cr^{2} -C^{2} r^{4} }{\, r\, \, (1+Cr^{2} )\, (2-Cr^{2} )}\frac{y'}{y}\right]\,e^{-\lambda } +\frac{3C^{2}r^{2} }{2\, (1+Cr^{2} )^{2} }  ;
\end{equation}

\section{Exact solution of the models for anisotropic stars}

The above equation Eq.(\ref{7}) is associated with the pressure anisotropy factor $\Delta$. First of all, our aim to construct the new anisotropy factor $\Delta $, which should be suitable to construct the models for anisotropic stars. That means $\Delta$ should be zero at the centre and positive for suitable choice of metric potential $y$. It is obvious that if $\Delta =0$, then $y=(1+Cr^{2} )^{3/2}$ is a particular solution of the equation Eq.(\ref{7}).

\noindent Now for constructing the expression for the pressure anisotropy, we suppose
\begin{equation}
\label{8}
y=(1+Cr^{2} +\beta \, Cr^{2} )^{3/2} ,  \beta>0 .
\end{equation}
is a particular solution of the Eq.(\ref{7}).
\begin{figure}[!htp]\centering
	\includegraphics[width=5.5cm]{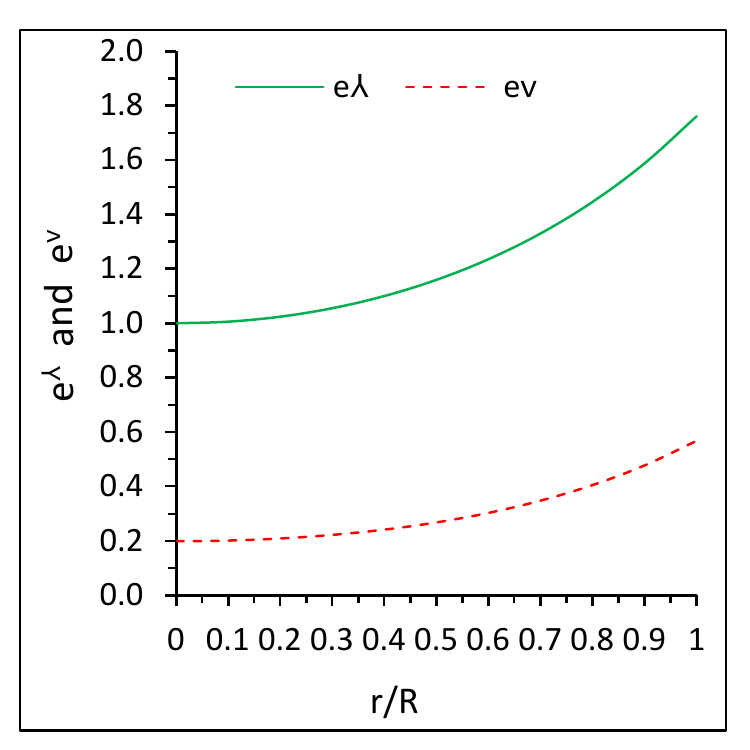}\includegraphics[width=5.5cm]{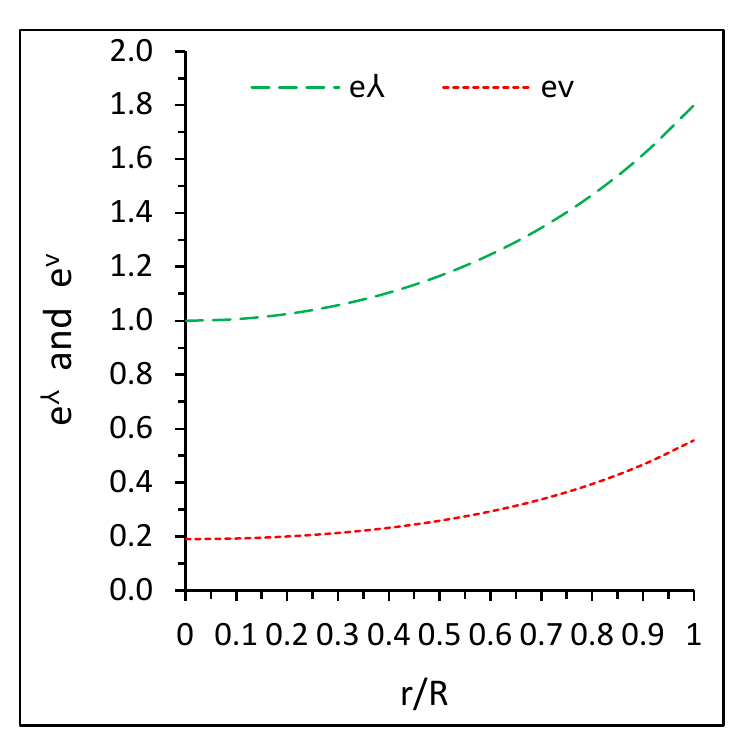}
		\caption{Variation of metric potentials $e^{\lambda } $ and $e^{\nu } $ with respect to fractional radius (r/R) for Her. X-1 (left panel) and for  RXJ 1856-37 (right panel).}
		\label{Fig1}
\end{figure}
 As we note  $y$ is a regular and monotonic increasing function [3]. Substituting $y$ from Eq.(\ref{8}) in Eq.(\ref{7}), we obtain:
\begin{equation}
\label{9}
\Delta =\frac{3C^{2}r^{2} \, [\, \beta +2\, \beta ^{2} -2\, \beta ^{2} \, Cr^{2} -2\beta \, Cr^{2} \, ]}{2 \, (1+Cr^{2} +\beta \, Cr^{2} )^{2} (1+Cr^{2} )^{2} }
\end{equation}
\begin{figure}[!htp]\centering
	\includegraphics[width=5.5cm]{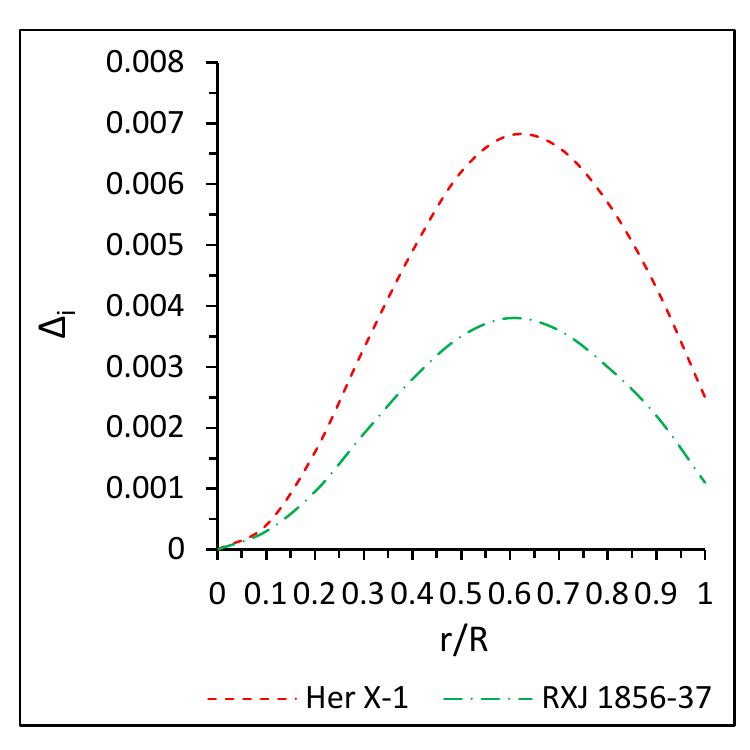}
	\caption{Variation of anisotropy factor ($\Delta_i=\frac{2\Delta}{C}$) with respect to fractional radius (r/R) for Her X-1 and RXJ 1856-37}
	\label{Fig2}
\end{figure}
 The above pressure anisotropy expression is finite and positive for
$0<Cr^{2} <\, \frac{2\, \beta +1}{2\,\beta +2}$ and $\beta >0$.

Our next aim to obtain the most general solution of the equation Eq.(\ref{7}) with anisotropic matter distribution, for this purpose we insert the expression of $\Delta $ in the Eq.(\ref{7}); we get
\begin{equation}
y''-\left[\frac{2+4\, Cr^{2} -C^{2} r^{4} }{\, r\, \, (1+Cr^{2} )\, (2-Cr^{2} )} \right]\, y'+\left[\frac{3C^{2}r^{2} [\, (\beta +1)^{2} \, C^{2} r^{4} +(2\, \beta ^{2} +4\beta \, +2)\, Cr^{2} \, -2\beta ^{2} -2\beta +1\, ]}{2\, (1+Cr^{2} )^{2} (1+Cr^{2} +\beta \, Cr^{2} )^{2} } \right]\, y=0
\label{10}
\end{equation}
Here Eq.\ref{10} is linear differential equation of second order. We shall apply the change of dependent variable method:
\noindent Let us the differential equation of the form:
\begin{equation}
\label{11}
Z''+S(r)\, y'+T(r)y=0,
\end{equation}
Suppose $z=z_{1} $ be the particular solution of the differential Eq.(\ref{11}), then the general solution will differential Eq.(\ref{11}) can be represented by $Z=z_{1} F$

\noindent where,
\begin{equation}
\label{12}
F=f_{1} +f_{2} \, \int _{}^{}\left\{\, e^{-\, [\, S(r)+(2z'/z)\, ]\, dr} \right\}\, dr  ;
\end{equation}
Now we suppose $y=(1+Cr^2+\beta \, Cr^{2} )^{3/2} =y_{1} $ is a particular solution of Eq.(\ref{10}), then most general solution of the differential Eq.(\ref{10}) is given by:
\begin{equation}
\label{13}
y=(1+Cr^{2} +\beta \, Cr^{2} )^{3/2} \, \left[\tilde{A}+\tilde{B}\int \exp \, \left\{\int \left(\frac{2+4\, Cr^{2} -C^{2} r^{4} }{\, r\, \, (1+Cr^{2} )\, (2-Cr^{2} )} -\frac{6(\beta +1)\, Cr^{2} }{r\, (1+Cr^{2} +\beta \, Cr^{2} )} \right)\, dr \right\}\, dr \right];
\end{equation}
After integration of above Eq.(\ref{13}),  we obtain
\begin{equation}
\label{14}
y=y_{1} \left[A+B\left\{\frac{9\, \{ \, 2w_{1} +w_{2} (1+Cr^{2} +\beta \, Cr^{2} )\} \sqrt{2+4\beta +2\beta ^{2} +(\beta +1)^{2} \, Cr^{2} \, (1-Cr^{2} )} }{8\, (1+Cr^{2} +\beta \, Cr^{2} )^{2} } +g(r)\right\}\right]
\end{equation}
where $A$ and $B$ are arbitrary constants of integration and
\begin{equation}
\label{15}
g(r)=\frac{\, w_{3} }{\sqrt{w_{4} } } \ln \left|\frac{\, \beta ^{2} Cr^{2} +(4\beta +3)\, Cr^{2} +4\beta ^{2} +7\beta +3)+\, 2\, \sqrt{w_{4} } \, \sqrt{2+4\beta +2\beta ^{2} +(\beta +1)^{2} \, Cr^{2} \, (1-Cr^{2} )} \, }{2\beta \, (2\beta +3)\, (1+Cr^{2} +\beta \, Cr^{2} )\, } \right|
\end{equation}
with $w_{1} =\frac{1\, }{\, (2\beta +3)}$, $w_{1} =\frac{1\, }{\, (2\beta +3)}$, $w_{3} =\frac{27}{16} \frac{(\beta ^{2} -2\beta -3)}{\beta  (2\beta +3)^{2} }$, and $w_{4} =\beta \, (2\beta +3)$.\\
The expressions for energy density, radial pressure and transverse pressure are obtained as:
\begin{equation}
\label{16}
\, \frac{\kappa \, c^{2} \rho }{C} =\frac{3\, \, (3+Cr^{2} )}{2\, (1+Cr^{2} )^{2} }
\end{equation}
\begin{figure}[!htp]\centering
	\includegraphics[width=5.5cm]{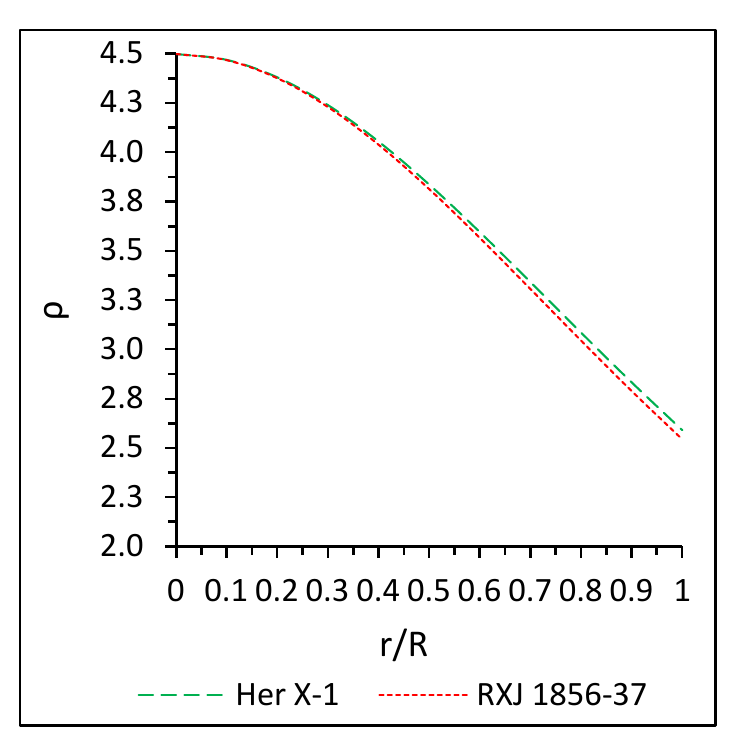}
	\caption{Variation of energy density with respect to fractional radius (r/R) for Her X-1 and RXJ 1856-37}
	\label{Fig3}
\end{figure}
\begin{equation}
\label{17}
\frac{\kappa \, p_{r} }{C} =\frac{1}{\, (1+Cr^{2} )} \left[\frac{(3\, \beta +3)\, \, (2-Cr^{2} )}{\, (1+Cr^{2} +\beta \, Cr^{2} )} -\frac{2(\, \beta +1)\, B\, (1+Cr^{2} )\, (2-Cr^{2} )}{\, y\, (1+Cr^{2} +\beta \, Cr^{2} )^{3/2} \, \sqrt{2+Cr^{2} -C^{2} r^{4} } } -\frac{3}{2} \right]
\end{equation}
\begin{figure}[!htp]\centering
	\includegraphics[width=5cm]{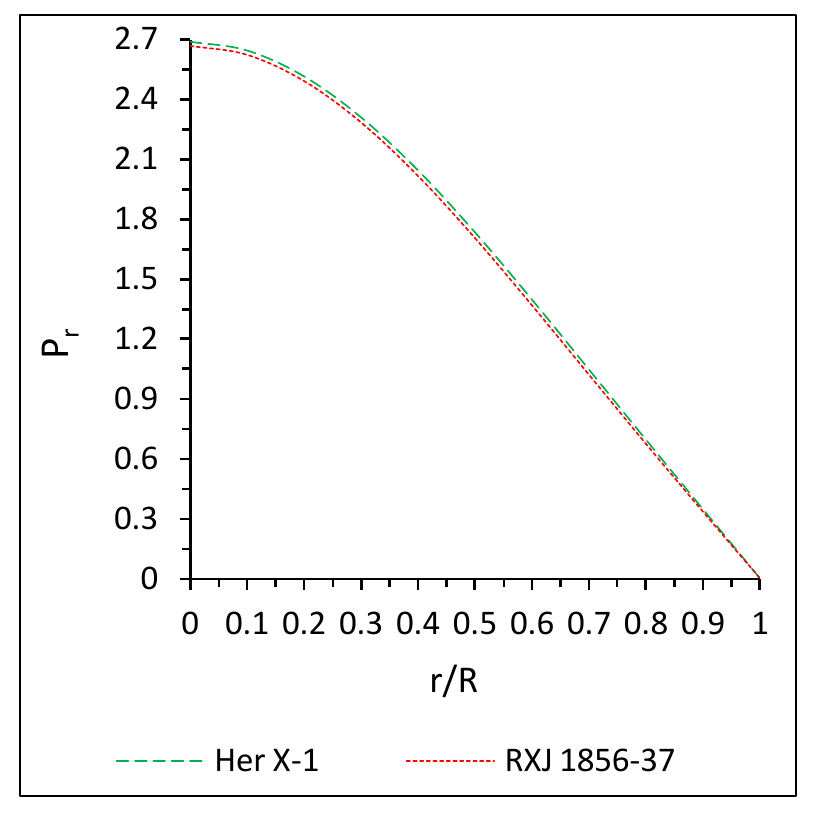}\includegraphics[width=5cm]{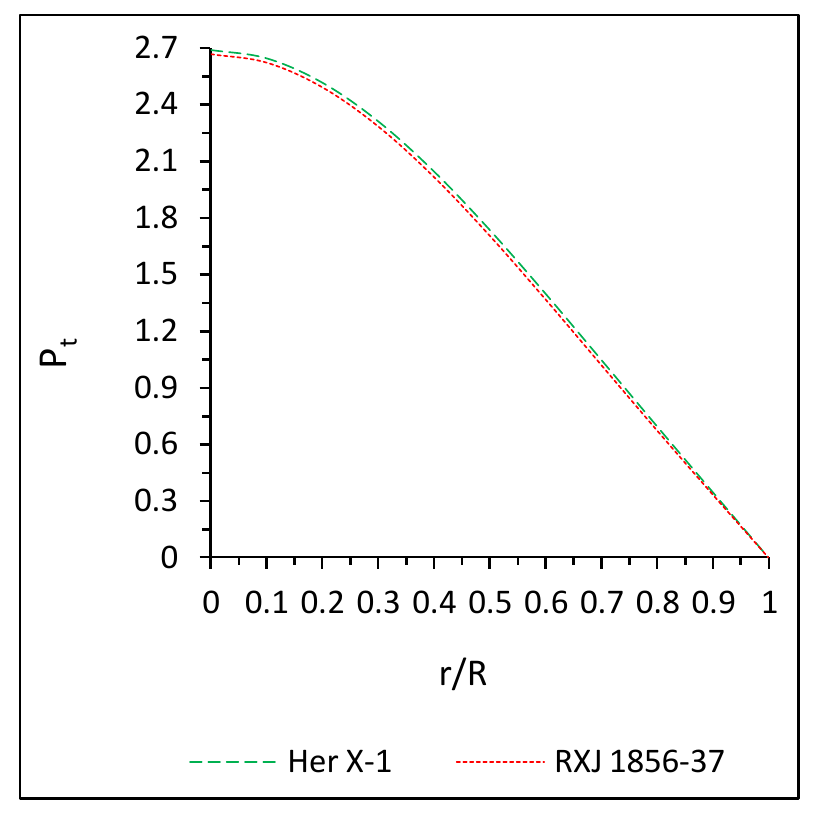}
	\caption{Variation of radial pressure (left panel) and transverse pressure (right panel) with respect to fractional
		radius (r/R) for Her X-1 and RXJ 1856-37.}
	\label{Fig4}
\end{figure}%
\begin{figure}[!htp]\centering
	\includegraphics[width=5cm]{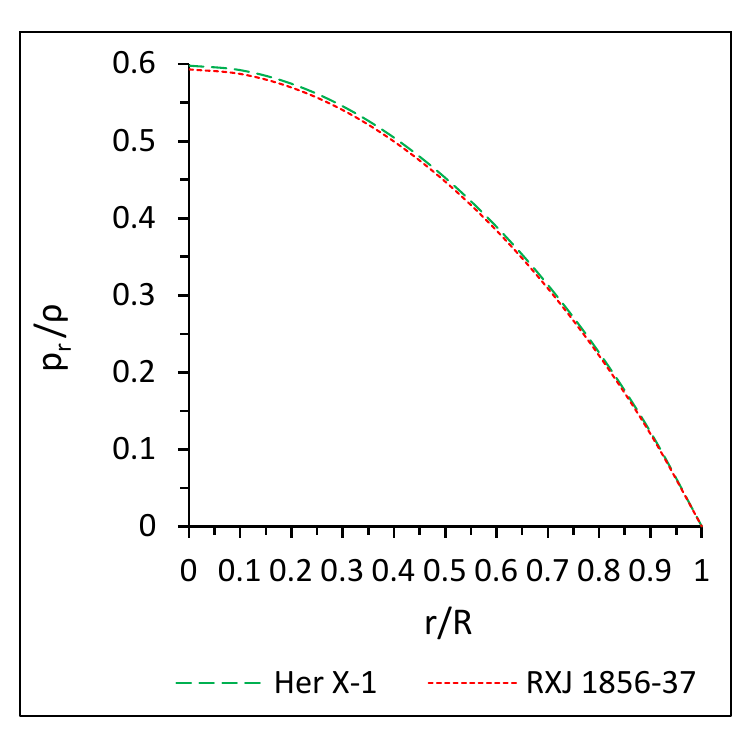}\includegraphics[width=5cm]{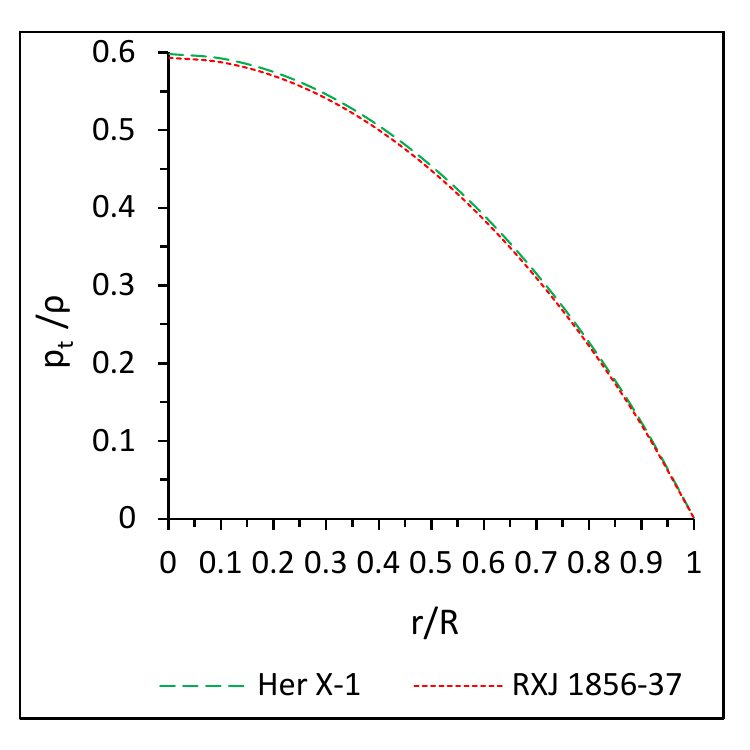}
	\caption{Variation of ratio of radial pressure verses density (left panel) and tangential pressure verses density (right panel) with respect to fractional radius (r/R) for Her X-1 and RXJ 1856-37.}
	\label{Fig5}
\end{figure}
\begin{eqnarray}
\label{18}
\frac{\kappa\ p_{t} }{C} =\frac{3\left( Cr^{2} +(\beta +1)(2-2Cr^{2} +C^{2} r^{4})-(\beta +1)^{2} Cr^{2}(C^{2} r^{4} +Cr^{2} -4)\right) }{(1+Cr^{2} +\beta Cr^{2})^{2} (1+Cr^{2} )^{2}}\nonumber\\ - \frac{2(\beta +1)B\sqrt{2-Cr^{2} }}{y(1+Cr^{2} +\beta Cr^{2} )^{3/2}\sqrt{1+Cr^{2}}} -\frac{3}{2(1+Cr^{2} )}
\end{eqnarray}
\section{Some physical features, matching condition and bounds on the parameters of the models}
\subsection{Regularity conditions at the centre:}
The radial pressure $p_r$, tangential pressure $p_t$ and density $\rho$ should be positive finite inside the star.
\noindent Density of star at centre:
\begin{equation} \label{19}
8\pi \, \rho _{0} =8\pi \, \rho (r=0)=\frac{9\, C}{2} ,
\end{equation}
 From Eq.(\ref{19})
\begin{equation} \label{20}
C=\frac{16\pi \, \rho _{0} }{9\, }
\end{equation}
Since density is positive and finite at centre. Then  $C$  is positive and finite.

Now the radial pressure at centre can be obtained from the Eq.(\ref{17}) as:
\begin{equation}
 \frac{p_{r} (r=0)}{C} =\left[6\beta -\frac{2\sqrt{2} (\beta +1)\, B}{y_{(r=0)} } +\frac{9}{2}
 \right]
  \label{21}
\end{equation}
where $p_{r} (r=0)>0$. Then, Eq.(\ref{21}) implies that:
\begin{equation}
\label{22}
\frac{A}{B} >\frac{4\sqrt{2} \, (\beta +1)}{12\beta +9} -\frac{9\, (2w_{1} +w_{2} )\, \sqrt{w_{4} +\beta +2} }{8} +\frac{w_{3} }{\sqrt{w_{4} } } \ln \left|\, \frac{2w_{4} +(\beta +3)+2\sqrt{w_{4} } \sqrt{w_{4} +\beta +3} }{2\, w_{4} } \right|
\end{equation}
\subsection{Matching Condition}
The exterior space--time of the star can be described by the Schwarzschild metric as
\begin{equation}
\label{23}
ds^{2} =\left(1-\frac{2M}{r} \right)\, dt^{2} -r^{2} (d\theta ^{2} +\sin ^{2} \theta \, d\phi ^{2} )-\left(1-\frac{2M}{r} \right)^{-1} dr^{2} ;
\end{equation}
Continuity of the metric coefficients $e^{\nu } $and $e^{\lambda }$ across the boundary surface of the models  $r=R$ between the interior and the exterior regions of the star yields the following conditions:
\begin{equation}
\label{24}
1-\frac{2M}{R} =y_{R}^{2}
\end{equation}
\begin{equation}
\label{25}
1-\frac{2M}{R} =e^{-\lambda (R)}  ,
\end{equation}
 where $y(r=R)=y_{R} $.

The radial pressure p${}_{r}$ =0 at $r=R$  gives
\begin{equation}
\label{26}
\frac{A}{B} =\frac{4\, (\beta +1)\, (1+CR^{2} )\, \sqrt{2+CR^{2} -C^{2} R^{4} } }{\psi _{R} \, \, [\, 6\, (\beta +1)\, (2+CR^{2} +C^{2} R^{4} )\, \psi _{R} \, -3\, (1+CR^{2} +\beta \, CR^{2} )^{2} (1+CR^{2} )\, ]} -\Omega (R)
\end{equation}
where,  $\psi _{R} =(1+CR^{2} +\beta \, CR^{2} )$
\begin{equation}
\label{27}
\Omega (R)=\frac{9\, \{ \, 2w_{1} +w_{2} (1+CR^{2} +\beta \, CR^{2} )\} \sqrt{2+4\beta +2\beta ^{2} +(\beta +1)^{2} \, CR^{2} \, (1-CR^{2} )} }{8\, (1+CR^{2} +\beta \, CR^{2} )^{2} } +G(R),
\end{equation}
\begin{equation}
\label{28}
G(R)=\frac{\, w_{3} }{\sqrt{w_{4} } } \ln \left|\frac{\, [\beta ^{2} CR^{2} +(4\beta +3)\, CR^{2} +4\beta ^{2} +7\beta +3]+\, 2\, \sqrt{w_{4} } \, \sqrt{2+4\beta +2\beta ^{2} +(\beta +1)^{2} \, CR^{2} \, (1-CR^{2} )} \, }{2\beta \, (2\beta +3)\, (1+CR^{2} +\beta \, CR^{2} )\, } \right|
\end{equation}
Eq.(\ref{24}) and Eq.(\ref{25}) respectively give:
\begin{equation}
\label{29}
B=\frac{\, \sqrt{2+CR^{2} -C^{2} R^{4} } }{\sqrt{2} \, \, (\psi _{R} )^{3/2} \, (1+CR^{2} )\, \left[\frac{A}{B} +\left\{\frac{9\, \{ \, 2w_{1} +w_{2} \, \psi _{R} \} \sqrt{2+4\beta +2\beta ^{2} +(\beta +1)^{2} \, CR^{2} \, (1-CR^{2} )} }{8\, (1+CR^{2} +\beta \, CR^{2} )^{2} } +G(R)\right\}\right]}
\end{equation}
\begin{equation} \label{30}
M=\frac{R}{2} \left[\frac{3\, CR^{2} )}{2\, (1+CR^{2} )} \right],
\end{equation}
\subsection{Causality conditions}
The speed of sound inside the star should be less than the speed of light i.e.
\[0\le V_{r} =\sqrt{\frac{dp_{r} }{c^{2} d\rho } } <1,    0\le V_{t} =\sqrt{\frac{dp_{t} }{c^{2} d\rho } } <1 ;  \]
\begin{figure}[!h]\centering
	\includegraphics[width=5cm]{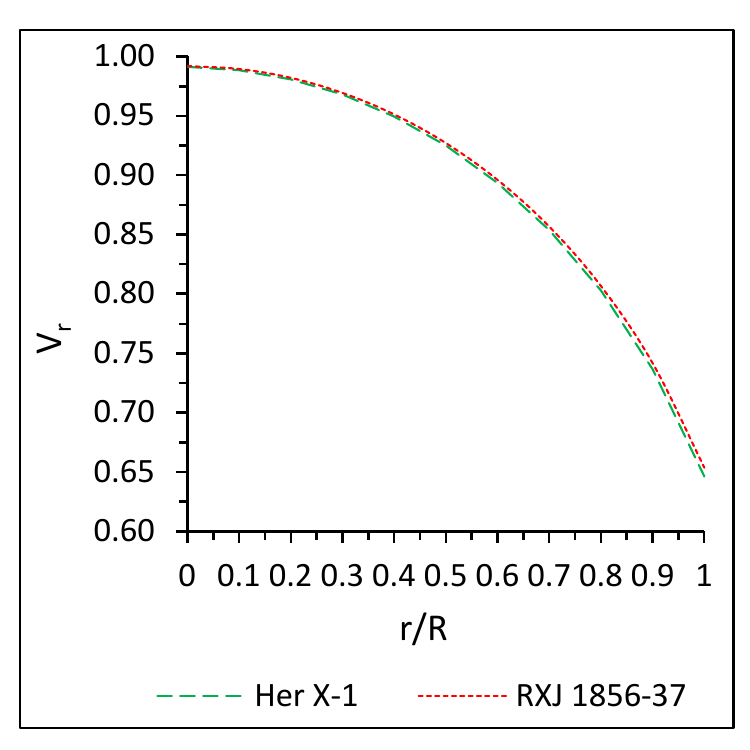}\includegraphics[width=5cm]{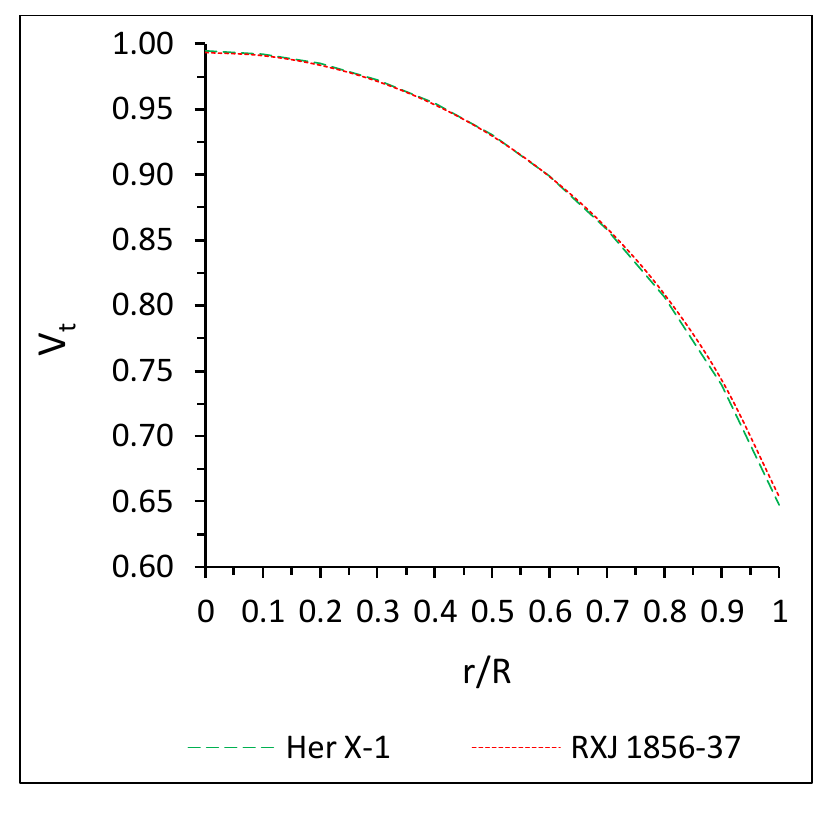}
	\caption{Variation of radial velocity (left panel) and transverse velocity (right panel) with respect to fractional radius (r/R) for Her X-1 and RXJ 1856-37.}
	\label{Fig6}
\end{figure}
\subsection{Energy conditions}
The anisotropic fluid sphere composed of fluid matter will satisfy the null energy condition, weak energy condition and strong energy condition, if the following inequalities hold simultaneously at all points in the star.

\noindent Null energy condition (NEC): $\rho \ge 0$

\noindent Weak energy condition (WEC${}_{r}$): $\rho +p_{r} \ge 0$

\noindent Weak energy condition (WEC${}_{t}$): $\rho +p_{t} \ge 0$

\noindent Strong energy condition (SEC): $\rho +p_{r} +2p_{t} \ge 0$
\begin{figure}[!h]\centering
	\includegraphics[width=5cm]{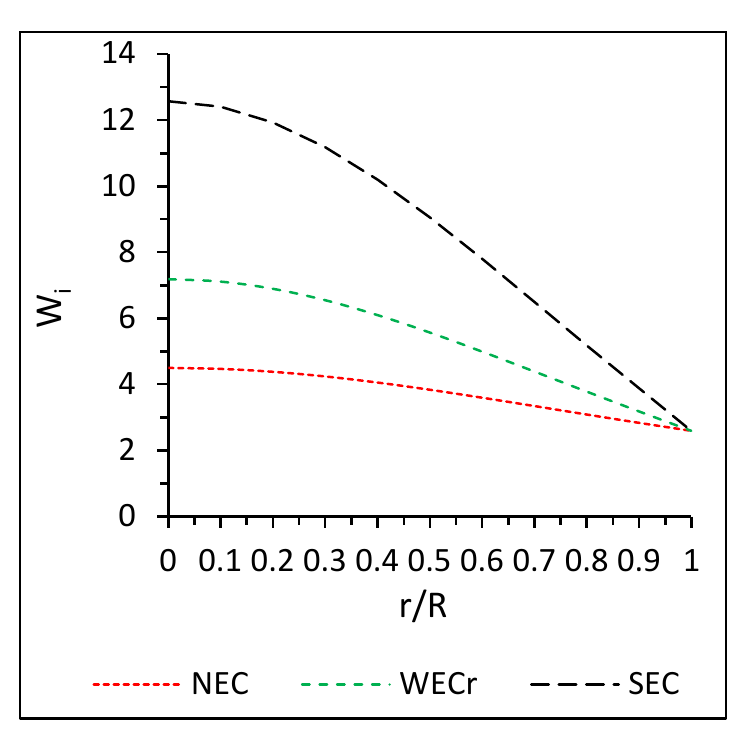}\includegraphics[width=5cm]{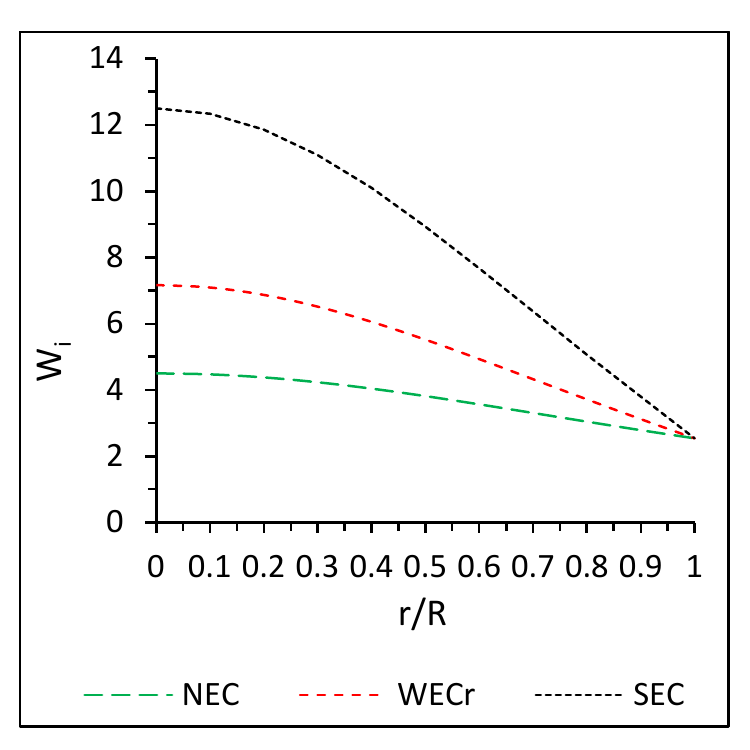}
	\caption{Variation of energy condition with respect to fractional radius (r/R) for Her X-1 (left panel) and RXJ 1856-37 (right panel).}
	\label{Fig7}
\end{figure}

\subsection{Generalized TOV equation: }
A star will remain in hydrostatic equilibrium provided that it's gravitational force which points inwards balances against its outward pointing internal pressures. Given a particular solution, the Tolman-Oppenheimer-Volkoff equation states a relationship between the mass, density and pressure which will allow the star to remain in hydrostatic equilibrium.

The generalized Tolman-Oppenheimer --Volkoff (TOV) equation
\begin{equation} \label{31}
-\frac{M_{G} (\rho +p_{r} )}{r^{2} } e^{\frac{\lambda -\nu }{2} } -\frac{dp_{r} }{dr} +\frac{2}{r} (p_{t} -p_{r} )=0;
\end{equation}
Where $M_{G} $ is the effective gravitational mass given by
\begin{equation} \label{32}
M_{G} (r)=\frac{1}{2} r^{2} e^{\frac{\nu -\lambda }{2} } \, \nu '
\end{equation}
The generalized Tolman-Oppenheimer --Volkoff (TOV) equation: TOV equations are used to calculate macroscopic features such as mass and radius of the star. The choice of the mass function is motivated by the fact that it gives a monotonically decreasing energy density in the stellar interior. Eq.(\ref{31}) is the modified form of TOV equation in the presence of anisotropy and it describes the equilibrium condition for an anisotropic fluid subject to gravitational ($F_{g} $), hydrostatic ($F_{h} $) and anisotropic stress ($F_{a} $) so that:
\begin{equation}
\label{33}
F_{g} +F_{h} +F_{a} =0 ,
\end{equation}
Where
\begin{equation} \label{34}
F_{g} =-\frac{1}{2} \nu '\, (\rho +p_{r} )
\end{equation}
\begin{equation} \label{35}
F_{h} =-\frac{dp_{r} }{dr}
\end{equation}
\begin{equation} \label{36}
F_{a} =\frac{2}{r} (p_{t} -p_{r} )
\end{equation}
The explicit form of above forces can be expressed in following forms:
\begin{equation}
\label{37}
F_{g} =-\frac{C^{2} r\, y'\, \, }{\, 8\pi \, y\, \, (1+Cr^{2} )} \left[\frac{3\, (\beta +1)\, \, (2-Cr^{2} )}{\, (1+Cr^{2} +\beta \, Cr^{2} )} -\frac{2\, (\beta +1)\, B\, (1+Cr^{2} )\, (2-Cr^{2} )}{\, y\, (1+Cr^{2} +\beta \, Cr^{2} )^{3/2} \, \sqrt{2+Cr^{2} -C^{2} r^{4} } } +\frac{3\, \, }{\, (1+Cr^{2} )} \right]
\end{equation}
\begin{equation}
\label{38}
F_{h} =-\frac{dp_{r} }{dr} =-\frac{C^{2} r}{8\pi}\,\frac{4\, (2+Cr^{2} -C^{2} r^{4} )\, \, [p_{1} (r)-p_{2} (r)-p_{3} (r)]-12\, (\beta +1)\, p_{4} \, (r)+3}{ \, (1+Cr^{2} )^{2} } ,
\end{equation}
\begin{equation}
\label{39}
F_{a} =\frac{2}{r} (p_{t} -p_{r} )=\frac{3C^{2}r\, [\, \beta +2\, \beta ^{2} -2\, \beta ^{2} \, Cr^{2} -2\beta \, Cr^{2} \, ]}{\,8\pi\, (1+Cr^{2} +\beta \, Cr^{2} )^{2} (1+Cr^{2} )^{2} }.
\end{equation}
\begin{figure}[!h]\centering
	\includegraphics[width=5cm]{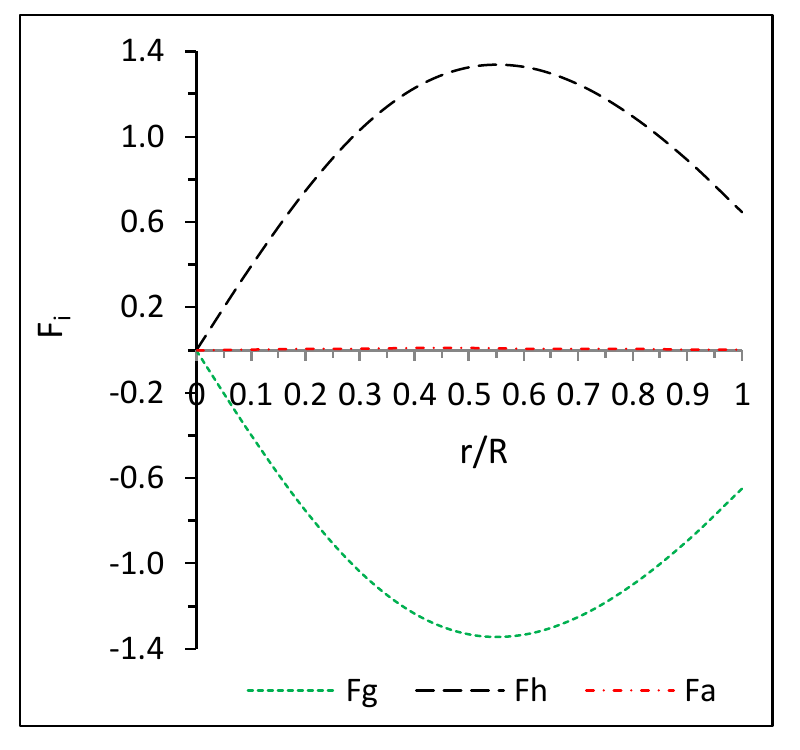}\includegraphics[width=5cm]{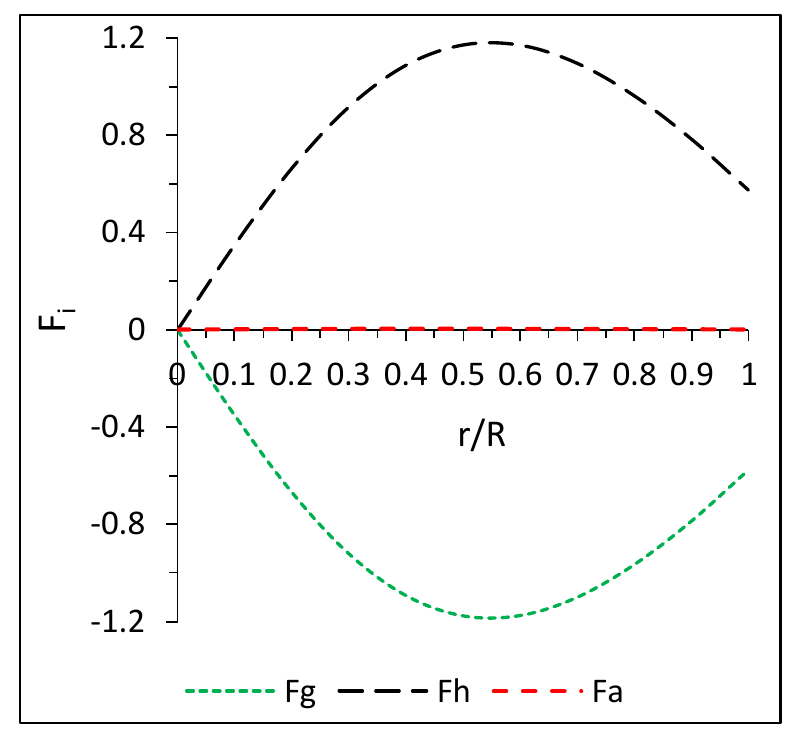}
	\caption{Variation of the different forces with respect to fractional radius (r/R) for Her X-1 (left panel) and RXJ 1856-37 (right panel) }
	\label{Fig8}
\end{figure}
\subsection{Stability conditions}
\subsubsection{Case-1:}
The study of dynamical stability of our solution is important; this requires computing the adiabatic index $\gamma$ and the pressure. According to the Le Chateliers principle (also known as local or microscopic stability condition) that the $p_r$ must be a monotonically non-decreasing function of $r$ such that $\frac{dp_r }{d\rho }\geq 0$ \cite{32}. Also, Heintzmann and Hillebrandt \cite{33} describe that neutron star with anisotropic equation of state is stable for $\gamma\ge 4/3$. This result says that star with $\gamma $= 4/3 marginally stable. If $\gamma $ is less than 4/3 then dynamical instability will occur, while if $\gamma $ is greater than 4/3 the relatively the star is stable. Thus for this case, positive anisotropy may slow down the growth if instability. We have seen that the presence of the anisotropic pressure in self-gravitating system can have dramatic effect on the dynamics and stability of the star. In particular, there are some novel features that are present only if the pressure is anisotropic, e.g., infinite core pressure, zero radial pressure and stable object with $\gamma\ge 4/3$. In the corresponding isotropic case stability immediately sets in if $\gamma $$<$ 4/3.
\begin{figure}[!h]\centering
	\includegraphics[width=5.5cm]{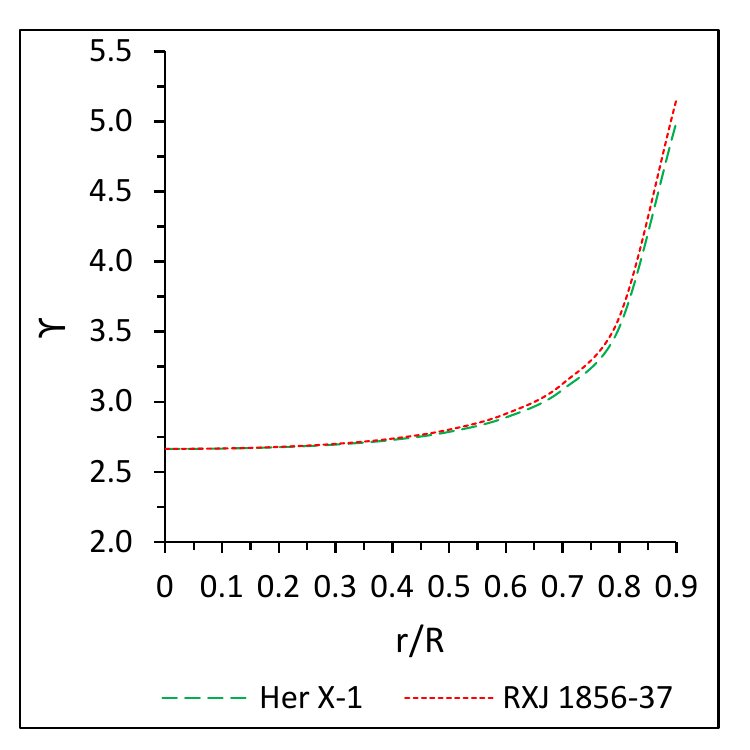}
	\caption{Variation of adiabatic index with respect to fractional radius (r/R) for Her X-1 and RXJ 1856-37}
	\label{Fig9}
\end{figure}
\subsubsection{Case-2:}
 For physically valid model, the velocity of sound should be less than or equal to velocity of light i.e. it should be within the range  $0\le V_{i}^{2} =(dp_{i} /c^{2} d\rho )\le 1$ \cite{34,35}.

The expressions for squares of velocities are given by
\begin{equation}
\label{40}
V_{r}^{2} =\left(\frac{4\, (2+Cr^{2} -C^{2} r^{4} )\, \, [p_{1} (r)-p_{2} (r)-p_{3} (r)]-12\, (\beta +1)\, p_{4} \, (r)+3}{-3\, (5+Cr^{2} )\,(1+Cr^{2})^{-1}} \right)
\end{equation}

\begin{equation}
\label{41}
V_{t}^{2} =\left(\frac{4\, (2+Cr^{2} -C^{2} r^{4} )\, \, [p_{1} (r)-p_{2} (r)-p_{3} (r)]-12(\beta +1)\, p_{4} \, (r)+3+p_{5} (r)}{-3\, (5+Cr^{2} )\,(1+Cr^{2})^{-1}} \right),
\end{equation}
where
\begin{equation}
\label{42}
p_{1} (r)=\frac{3\, (\beta +1)^{2} }{4\, (1+Cr^{2} +\beta \, Cr^{2} )^{2} }
\end{equation}
\begin{equation}
\label{43}
p_{2} (r)=\frac{\, (\beta +1)^{2} \, [3y\, \sqrt{1+Cr^{2} +\beta \, Cr^{2} } \sqrt{2+Cr^{2} -C^{2} r^{4} } -2B(1+Cr^{2} )]^{2} }{4y^{2} (1+Cr^{2} +\beta \, Cr^{2} )^{3} \, (2+Cr^{2} -C^{2} r^{4} )} ,
\end{equation}
\begin{equation}
\label{44}
p_{3} (r)=\frac{3\, (\beta +1)\, B(1+Cr^{2} )}{2y\, (1+Cr^{2} +\beta \, Cr^{2} )^{3/2} (2+Cr^{2} -C^{2} r^{4} )^{3/2} } ,
\end{equation}
\begin{equation}
\label{45}
p_{4} (r)=\frac{3\, \, }{2\, (1+Cr^{2} +\beta \, Cr^{2} )} -\frac{\, B\, (1+Cr^{2} )}{\, y\, (1+Cr^{2} +\beta \, Cr^{2} )^{3/2} \, \sqrt{2+Cr^{2} -C^{2} r^{4} } } ,
\end{equation}
\begin{equation}
\label{46}
p_{5} (r)=\, \frac{3\, \beta \, \, [\, 2\beta ^{2} Cr^{2} (2C^{2} r^{4} -3Cr^{2} -1)+\beta (8C^{3} r^{6} -9 C^{2} r^{4} -9Cr^2+2)+4C^{3} r^{6} -3C^{2} r^{4} -6Cr^2+1\,]}{(1+Cr^{2} +\beta \, Cr^{2} )^{3} \, (1+Cr^{2} )}  .
\end{equation}
\begin{figure}[!h]\centering
	\includegraphics[width=5cm]{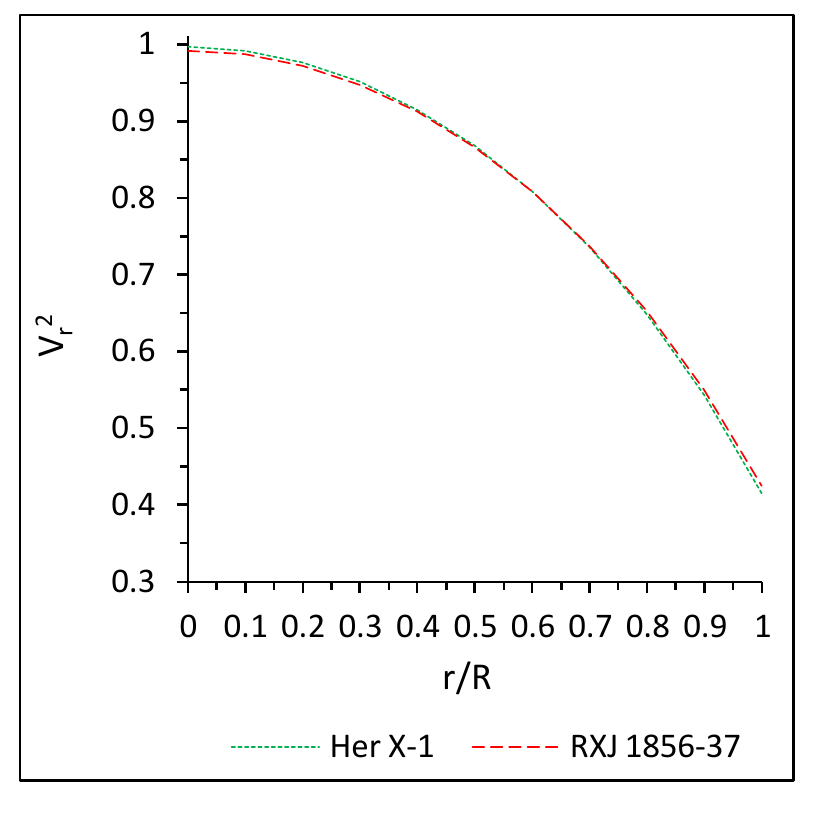}\includegraphics[width=5cm]{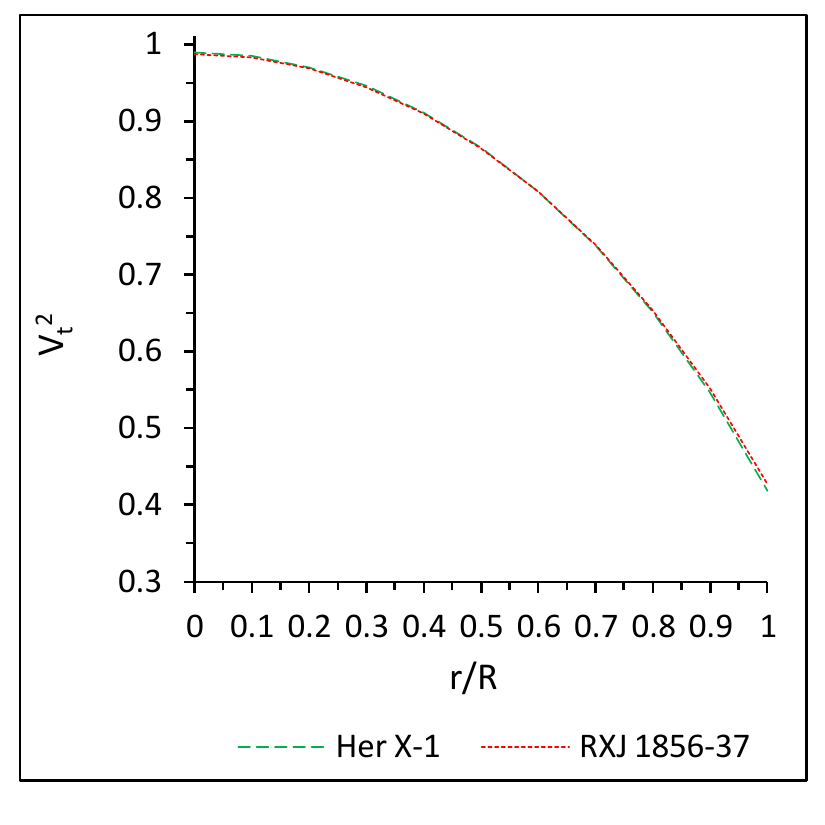}
	\caption{Variation of square of radial velocity (left panel)and transverse velocity (right panel) with respect to fractional radius (r/R) for Her X-1 and RXJ 1856-37}
	\label{Fig10}
\end{figure}
Form Fig. 10, it is clear that the radial and transverse velocity of sound is satisfying the inequalities $0\le V_{i}^{2} =(dp_{i} /c^{2} d\rho )\le 1$ everywhere inside the star i.e. $0\le V_{r}^{2} \le 1$ and $0\le V_{t}^{2} \le 1$.

 In order to determine the stability of anisotropic star, Herrera's [34] proposed the cracking (also known as overturning) concept which states that the region is potentially stable in which radial speed of sound is greater than the transverse speed of sound.

For this purpose, we calculate the difference of velocities which can be expressed as:

\begin{equation}
\label{47}
V_{\, t}^{2} -V_{\, r}^{2} =\, \frac{ \beta \,  [\, 2\beta ^{2} Cr^{2} (2C^{2} r^{4} -3Cr^{2} -1)+\beta (8C^{3}r^{6}-9 C^{2} r^{4}-9Cr^2+2)+4C^{3} r^{6}-3C^{2} r^{4}-6Cr^2+1]}{-(1+Cr^{2} +\beta \, Cr^{2} )^{3} \, (5+Cr^{2} )}.
\end{equation}
\begin{figure}[!h]\centering
	\includegraphics[width=5.5cm]{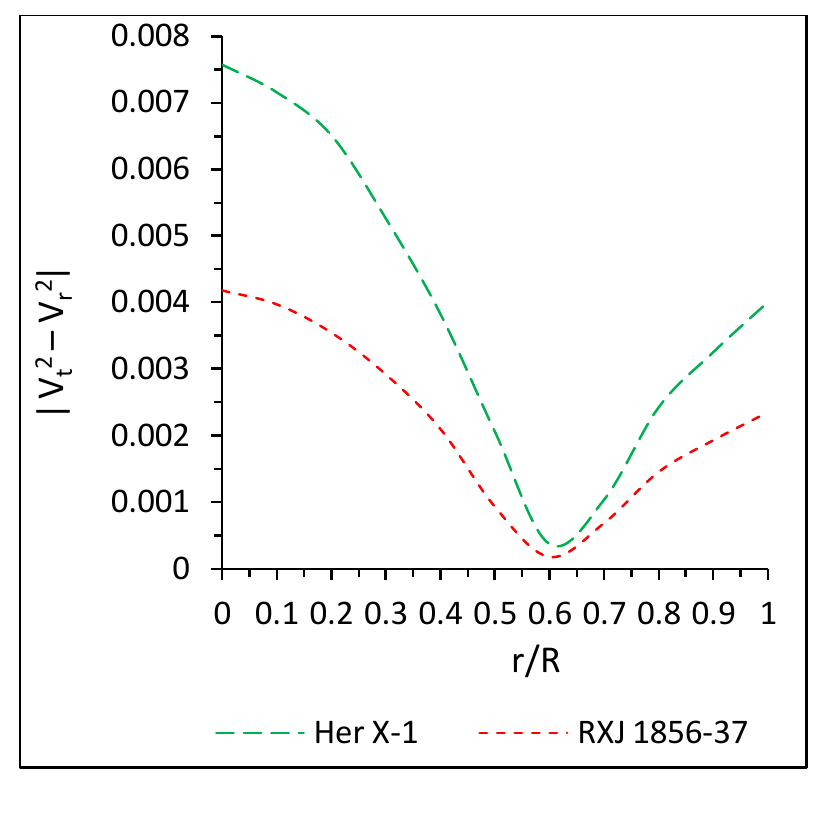}
	\caption{Variation of the absolute value of square of sound velocity with respect to fractional radius (r/R)}
	\label{Fig11}
\end{figure}
 It is observed that $\left|V_{\, t}^{2} -V_{\, r}^{2} \right|$ at the centre also lies between 0 and 1 (Fig.11). This implies that we must have $0<\frac{\beta \, (2\beta +1)}{5} \le 1$. Then we conclude that $\beta $ should satisfy the following inequality $0<\beta \le \frac{-1+\sqrt{41} }{4} $.

\section{Effective mass-radius relation and surface red-shift of compact star:}
Let us now we will discuss about the effective mass-radius relation of the compact star. Buchdahl \cite{1} has given an absolute bound on the maximum allowable mass-radius ratio (M/R) for spherically symmetric perfect fluid spheres which is $2M/R\le 8/9$ (in the unit, c = G = 1). This upper bound of mass to radius ratio shows that for a given radius a static perfect fluid sphere cannot be arbitrarily massive. However, Mak and Harko \cite{17} have given a more generalized expression for the mass-to-radius ratio.
\begin{figure}[!h]\centering
	\includegraphics[width=5cm]{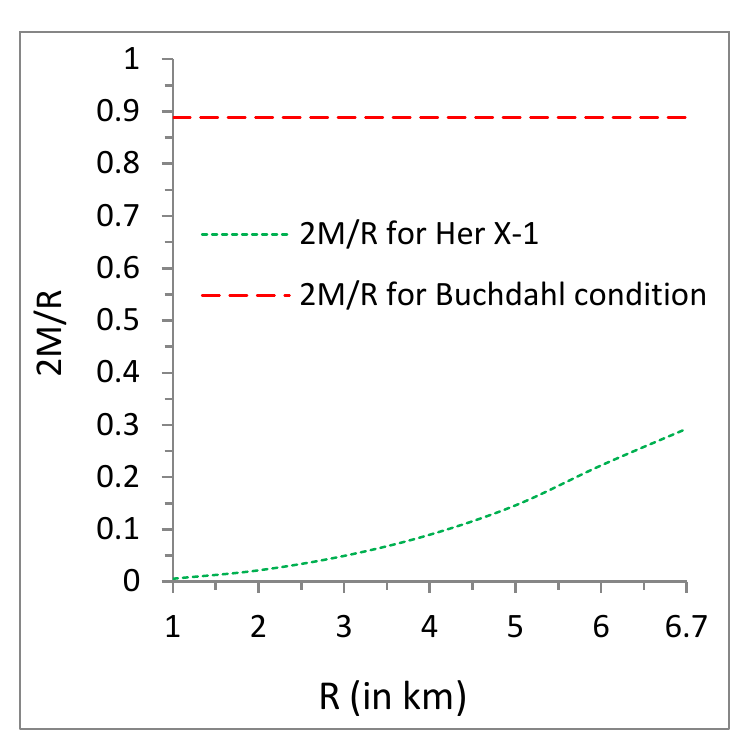}\includegraphics[width=5cm]{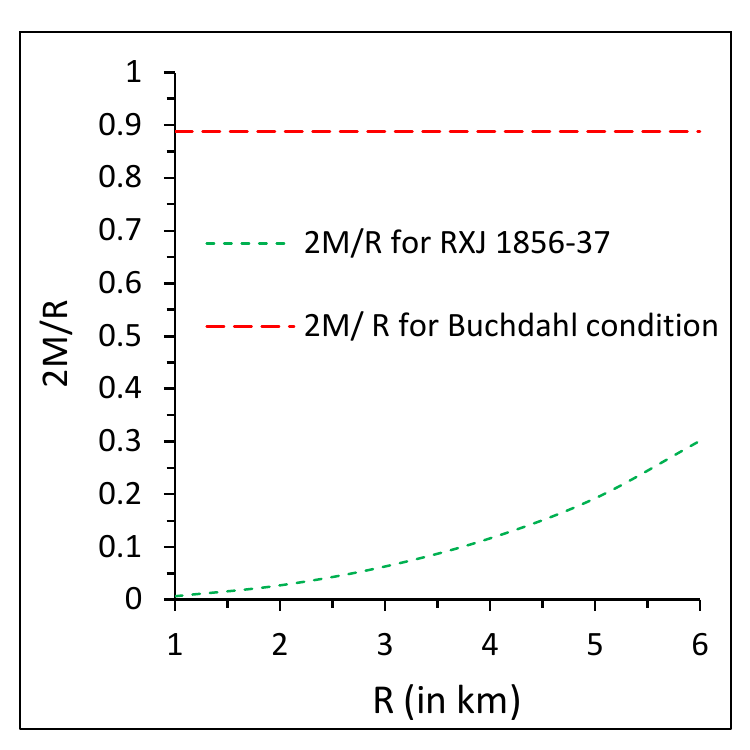}
	\caption{Variation of 2M/R with respect radius (R) for Her X-1 (left panel) and for RXJ 1837-56 (right
		panel)
	}
	\label{Fig12}
\end{figure}
In our present compact star models, the effective gravitational mass has of the form as
\begin{equation}
\label{48}
M_{eff} =4\pi \int _{0}^{R}\rho \,  r^{2} dr=\frac{1}{2} R\, [\, 1-e^{-\lambda (R)} ]=\frac{1}{2} R\, \left[\frac{3CR^{2} }{2\, (1+CR^{2} )} \right],
\end{equation}
Therefore the compactness of the star can be expressed as
\begin{equation} \label{49}
u=\frac{M_{eff} }{R} =\frac{1}{2} \, \left[\frac{3CR^{2} }{2\, (1+CR^{2} )} \right];
\end{equation}
From Fig.(\ref{12}), we note that mass- radius ratio (2M/R) of our anisotropic fluid models are satisfying the Buchdahl condition  i.e. $2M/R\le 8/9=0.88889$.

Then the corresponding to the above compactness ($u$), the surface red-shift ($Z_{s} $) is obtained as
\begin{equation}
\label{50}
Z_{s} =(1-2u)^{-\frac{1}{2} } \, -1=e^{\frac{1}{2} \lambda (R)} -1=\sqrt{\frac{2(1+CR^{2} )}{2-CR^{2} } } \, -1.
\end{equation}
\begin{figure}[!h]\centering
	\includegraphics[width=5.5cm]{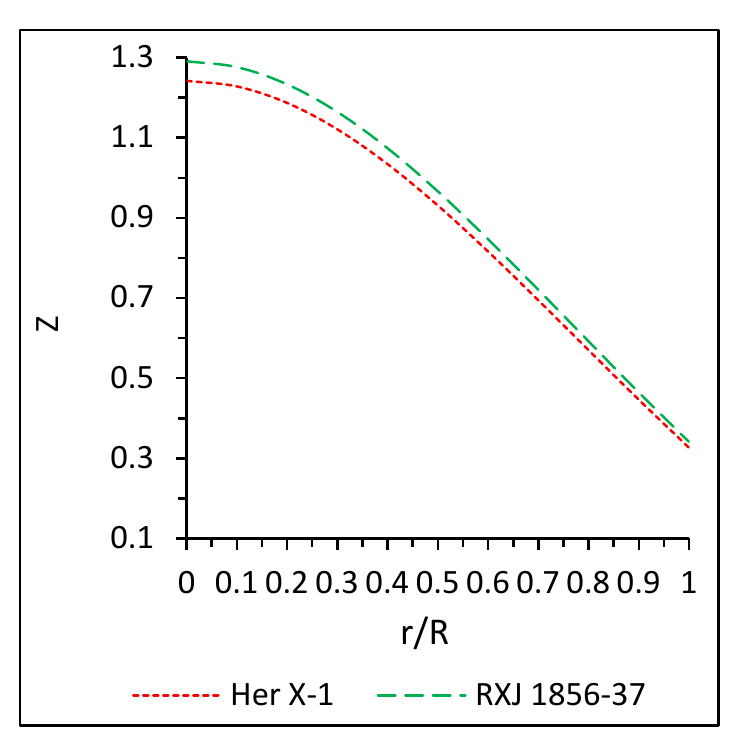}
	\caption{Variation of red-shift with respect to fractional radius for Her X-1 and RXJ 1856-37}
	\label{Fig13}
\end{figure}
\section{Model parameters and comparison with some of the observational compact stars}

In the present section, we prepare several data sheets of the physical parameters for the models in the Tables 1-4. Also we compared those with some of the compact objects, i.e. compact star Her X-1 and RXJ 1856-37 in table 4. We have already plotted the graphs for those data points in the previous section 3 and 4. (See Figs. 1-13).  In our present observation we obtained a stable compact object model with physical parameters $R = 6.6963$ Km and Mass $M = 0.9805M_{\odot}$ for Her X-1 while $M = 0.9031M_{\odot}$, $R = 6$ Km for RXJ 1856-37. However the surface redshift turns out to be Z=0.3268 for Her X-1 and Z=0.3411 for RXJ 1856-37, which shows that it is lying within the range $Z\le 2\, $[1, 36, 37] and $0<Z\le 1$[38-42] Also the summary of proposed tables as follows: Tables 1-2 have data for the physical parameters for compact star Her X-1 and RXJ 1856-37. In Table 3, the values of mass and radius are given for different constants A, B and C. Under these data values, we obtained some physical quantities of the compact star like central energy density, central pressure and surface energy density in Table 4. Under these observations we conclude that our data values are quite satisfactory for compact star. It can represent the observational compact star Her X-1 with central energy density 2.1793$\times 10^{15} $ $gm/cm^{3} $ and RXJ 1856-37 with central energy density 2.8226 $\times 10^{15} $$gm/cm^{3} $.

\begin{table}[h]
	\centering
	\caption{Values of different physical parameters of Her X-1 for$\beta $= 0.035, $C\, R^{2} $ = 0.4044, $M= 0.9805M_{\odot}$, R = 6.6963 Km and RXJ 1856-37 for$\beta $= 0.02, $C\, R^{2} $ = 0.4205, $M = 0.9031M_{\odot}$, R = 6 Km.}
	\label{Table1}
		\begin{tabular}{cccccc|ccccc}\hline
		\multicolumn{6}{c|}{\textbf{Her  X-1}} & \multicolumn{5}{|c}{\textbf{RXJ 1856-37}} \\ \hline
		$r/R$ & $P_{r}$ & $P_{t}$ & $D$ & $V_{r}$ & $V_{t}$ & $P_{r}$ & $P_{t}$ & $D$ & $V_{r}$ & $V_{t}$ \\ \hline
		0.0 & 2.6907 & 2.6907 & 4.5000 & 0.9985 & 0.9947 & 2.6683 & 2.6683 & 4.5000 & 0.9959 & 0.9938 \\ \hline
		0.1 & 2.6464 & 2.6466 & 4.4698 & 0.9959 & 0.9923 & 2.6233 & 2.6234 & 4.4686 & 0.9934 & 0.9914 \\ \hline
		0.2 & 2.5167 & 2.5175 & 4.3814 & 0.9883 & 0.9850 & 2.4920 & 2.4925 & 4.3768 & 0.9859 & 0.9841 \\ \hline
		0.3 & 2.3115 & 2.3132 & 4.2403 & 0.9752 & 0.9725 & 2.2847 & 2.2856 & 4.2305 & 0.9732 & 0.9717 \\ \hline
		0.4 & 2.0453 & 2.0477 & 4.0553 & 0.9565 & 0.9545 & 2.0164 & 2.0178 & 4.0391 & 0.9549 & 0.9538 \\ \hline
		0.5 & 1.7348 & 1.7379 & 3.8367 & 0.9314 & 0.9303 & 1.7049 & 1.7067 & 3.8137 & 0.9304 & 0.9299 \\ \hline
		0.6 & 1.3971 & 1.4004 & 3.5953 & 0.8991 & 0.8989 & 1.3677 & 1.3696 & 3.5658 & 0.8989 & 0.8988 \\ \hline
		0.7 & 1.0467 & 1.0500 & 3.3417 & 0.8580 & 0.8586 & 1.0202 & 1.0220 & 3.3062 & 0.8589 & 0.8593 \\ \hline
		0.8 & 0.6947 & 0.6975 & 3.0848 & 0.8055 & 0.8070 & 0.6736 & 0.6751 & 3.0445 & 0.8079 & 0.8070 \\ \hline
		0.9 & 0.3465 & 0.3486 & 2.8321 & 0.7372 & 0.7394 & 0.3339 & 0.3350 & 2.7881 & 0.7417 & 0.7430 \\ \hline
		1.0 & 0.0000 & 0.0012 & 2.5891 & 0.6440 & 0.6471 & 0.0000 & 0.0006 & 2.5427 & 0.6520 & 0.6538 \\ \hline
	\end{tabular}
\end{table}
\begin{table}
	\centering
	\caption{Values of different physical parameters of Her X-1 for$\beta $= 0.035, $C\, R^{2} $ = 0.4044, $M = 0.9805M_{\odot}$, R = 6.6963 Km and RXJ 1856-37 for $\beta$= 0.02, $C\, R^{2} $ = 0.4205, $M = 0.9031M_{\odot}$, R = 6 Km.}
	\label{Table2}
	\begin{tabular}{cccccc|ccccc} \hline
		\multicolumn{6}{c|}{\textbf{Her  X-1}} & \multicolumn{5}{|c}{\textbf{RXJ 1856-37}} \\ \hline
		$r/R$ & $e^{\lambda } $ & $e^{\nu } $ & $\gamma $ & $Z$ & $\Delta_i $ & $e^{\lambda } $ & $e^{\nu } $ & $\gamma $ & $Z$ & $\Delta_i $ \\ \hline
		0.0 & 1 & 0.1990 & 2.6644 & 1.2416 & 0.0000 & 1 & 0.1906 & 2.6645 & 1.2908 & 0.0000 \\ \hline
		0.1 & 1.0061 & 0.2015 & 2.6673 & 1.2276 & 0.0004 & 1.0063 & 0.1930 & 2.6680 & 1.2761 & 0.0003 \\ \hline
		0.2 & 1.0245 & 0.2092 & 2.6769 & 1.1865 & 0.0016 & 1.0254 & 0.2005 & 2.6792 & 1.2331 & 0.00095 \\ \hline
		0.3 & 1.0556 & 0.2224 & 2.6956 & 1.1207 & 0.0033 & 1.0579 & 0.2135 & 2.7007 & 1.1643 & 0.0019 \\ \hline
		0.4 & 1.1003 & 0.2417 & 2.7287 & 1.0339 & 0.0049 & 1.1044 & 0.2325 & 2.7381 & 1.0737 & 0.0028 \\ \hline
		0.5 & 1.1597 & 0.2682 & 2.7861 & 0.9308 & 0.0062 & 1.1664 & 0.2587 & 2.8020 & 0.9662 & 0.0035 \\ \hline
		0.6 & 1.2355 & 0.3032 & 2.8888 & 0.8161 & 0.0068 & 1.2457 & 0.2932 & 2.9146 & 0.8469 & 0.0038 \\ \hline
		0.7 & 1.3299 & 0.3483 & 3.0865 & 0.6945 & 0.0066 & 1.3446 & 0.3377 & 3.1282 & 0.7207 & 0.0036 \\ \hline
		0.8 & 1.4459 & 0.4056 & 3.5304 & 0.5702 & 0.0057 & 1.4664 & 0.3945 & 3.6027 & 0.5921 & 0.0030 \\ \hline
		0.9 & 1.5876 & 0.4778 & 4.9853 & 0.4467 & 0.0043 & 1.6158 & 0.4662 & 5.1434 & 0.4646 & 0.0022 \\ \hline
		1.0 & 1.7603 & 0.5681 & $\infty $ & 0.3268 & 0.0025 & 1.7987 & 0.5560 & $\infty $ & 0.3411 & 0.0011 \\ \hline
	\end{tabular}
\end{table}
\begin{table}
	\centering
	\caption{Values of the model parameters A, B, C and $\beta$ for different compact stars:}
	\label{Table3}
\begin{tabular}{@{}lrrrrrrr@{}} \hline
Compact star\\ candidates & $M(M_{\odot})$ & $R$(Km) & $M/R$ & $\beta $ & $A$ & $B$ & C \\ \hline
Her. X-1  & 0.9805 & 6.7 & 0.216 & 0.035 & 127.0022 & 0.8188 & 9.0185$\times 10^{-13} $ \\ \hline
RXJ 1856-37 & 0.9031 & 6 & 0.222 & 0.020 & 362.1375 & 0.7846 & 1.1680$\times 10^{-12} $ \\ \hline
\end{tabular}
\end{table}
\begin{table}
	\centering
	\caption{Energy densities, central pressure and Buchdahl condition for different compact star candidates for the above parameter values of Tables 1 - 3}
	\label{Table4}
\begin{tabular}{@{}lrrrr@{}} \hline
Compact star & Central Density & Surface density & Central pressure & Buchdahl  \\
candidates & $gm/cm^{3} $ & $gm/cm^{3}$ & $dyne/cm^{2} $ & condition\\\hline
Her. X-1  & 2.1793$\times 10^{15} $ & 1.2539$\times 10^{15} $ & 1.1730$\times 10^{36} $ &$2M/R\le 8/9=0.88889$  \\ \hline
RXJ 1856-37 & 2.8226$\times 10^{15} $ & 1.5949$\times 10^{15} $ & 1.5066$\times 10^{36}$ & $2M/R\le 8/9=0.88889$  \\ \hline
\end{tabular}
\end{table}
\section{Discussions and Conclusion:}

In this paper, we constructed the general anisotropic solution of well know Buchdahl perfect fluid solution for relativistic compact objects that is globally neutral anisotropic and satisfies the conditions of hydrostatic equilibrium. In particular, our results are dependent of the matter distribution and other physical parameters. As an important step, first we have started with $e^{\lambda} $form Buchdahl metric \cite{1} and $e^{\nu } $through lake \cite{3} assumption to construct the anisotropy factor $\Delta $, which is physically valid.  After that we have obtained the general solution of anisotropic fluid distribution. In the next section, we joined smoothly the interior metric Eq.(\ref{1}) with the Schwarzschild metric on the boundary of the star (r=R) and calculated the arbitrary constants A and B.  Also we have checked all regularity and stability conditions of compact stars with anisotropic matter distribution which are quite satisfied and also helpful in gravitational description of bodies such as HerX-1 \& RXJ1856-37.

As a detailed discussion, we will explore our physical results as follows:

(A) Regularity, Causality and well behaved conditions:  The metric potential $e^{\lambda } $and $e^{\nu } $ are regular at the centre and monotonic increasing (Fig.1), (ii) energy density $\rho $,  radial pressure $p_{r} $ and tangential pressure $p_{t} $ are positive and finite inside the star (Fig. 3 and 4). Also, the ratio of pressure versus density is monotonically decreasing away from the center (Fig.5). (iii) Fig.6 shows that the velocity of sound is less than the velocity of light inside the fluid sphere as well as it is monotonically decreasing away from the centre i.e. well behaved.

\noindent (B) Energy and stability conditions: (i) From Fig. 7, we conclude that our models satisfy the all energy conditions at each points inside the star. (ii) According to the Heintzmann and Hillebrandt \cite{30}, the stable neutron star with anisotropic equation of state has an adiabatic index $\gamma $ $>$ 4/3 as which can be observed from Fig.(\ref{9}) and Table 2 of our proposed models.  (iii) Also, we note that the velocity of sound should be within the range i.e. $0<V_{i}^{2} \le 1$ [34,35] and this can be seen from Fig. (\ref{10}).

\noindent (C) Generalized TOV equation: The plot for generalized Tolman-Openheimer-Volkoff equations is given by the Fig.(\ref{8}). We conclude from this figure that the system is counter balance under the different forces, i.e. the gravitational force, hydrostatic force and anisotropic stress, and system attains a static equilibrium. However, the gravitational force is dominating the hydrostatic force and it is balanced by the joint action of hydrostatic force and anisotropic stress, where the anisotropic stress has a negligible role to the action of equilibrium condition. These physical features represents that the models are stable.

\noindent (D) Anisotropy features and red shift: (i) the behavior of anisotropy is shown in Fig. Eq.(\ref{2}). We observe form this figure that the anisotropy is increasing with increase the radius and it attains maximum value inside the star. After reaching maximum it start decreasing with radial distance as such behavior also shown by MaK and Harko [43, 44]. However, the anisotropy is vanishing at the centre of the star. (ii) As we can see from table 1, the red shift of the compact star Her X-1 and RXJ 1856-37 is maximum at centre and minimum at the surface and corresponding values as follows: (a) at the center Z= 1.2416 and at the surface  Z=0.3268 for Her X-1, (b)  at centre  Z= 1.2908 and at surface Z=0.3411. Also, the behavior of red shift inside the compact star is shown by the Fig.(\ref{13}).
\section*{acknowledgment}
Dr. S.K.Maurya \textit{et.al.} acknowledge continuous support and encouragements from University of
Nizwa administrations.

\end{document}